\documentclass[twocolumn,twoside]{IEEEtran}

\ifCLASSINFOpdf

\else

\fi

\ifCLASSOPTIONcompsoc
\usepackage[caption=false, font=normalsize, labelfont=sf, textfont=sf]{subfig}
\else
\usepackage[caption=false, font=normalsize]{subfig}
\fi
\usepackage{lipsum}%
\usepackage[dvipsnames]{xcolor}

\usepackage{balance}
\usepackage{multicol}   
\usepackage{cite}
\usepackage{gensymb}
\usepackage{multirow}
\usepackage{graphics}
\usepackage{epsfig}
\usepackage{graphicx}
\usepackage{epstopdf}
\usepackage{textcomp}
\usepackage{amsmath}
\usepackage{mathtools}
\usepackage{filecontents}
\usepackage{lipsum,color}
\usepackage{amssymb}
\usepackage{float}
\usepackage{comment}

\usepackage{amsthm}

\usepackage{braket}
\usepackage{times} 
\usepackage{amsthm}  

\usepackage{amsfonts}

\theoremstyle{break}

\theoremstyle{break}
\begin{document}
\title{AI-Assisted Next-Gen Outdoor Optical Networks: Camera Sensing for Monitoring and User Localization}

\author{Meysam~Ghanbari,~Mohammad~Taghi~Dabiri,
	~Rula~Ammuri, \\
	~Mazen~Hasna,~{\it Senior Member,~IEEE}, 
	~and~Khalid~Qaraqe,~{\it Senior Member,~IEEE}
\thanks{M. Ghanbari and M.T. Dabiri are with the Qatar Center for Quantum Computing, College of Science and Engineering, Hamad Bin Khalifa University, Doha, Qatar. email: (megh89467@hbku.edu.qa; mdabiri@hbku.edu.qa).}
	
	\thanks{Rula Ammuri is with Professionals for Smart Technology (PST), Amman, Jordan (email: rammuri@pst.jo).}
	
	\thanks{Mazen Hasna is with the Department of Electrical Engineering, Qatar University, Doha, Qatar (email: hasna@qu.edu.qa).}
	
	\thanks{Khalid A. Qaraqe is with the College of Science and Engineering, Hamad Bin Khalifa University, Doha, Qatar, and also with the Department of Electrical Engineering, Texas A\&M University at Qatar, Doha, Qatar (email: kqaraqe@hbku.edu.qa).}
	\thanks{This publication was made possible by NPRP14C-0909-210008 from the Qatar Research, Development and Innovation (QRDI) Fund (a member of Qatar Foundation), Texas A\&M University at Qatar, and Hamad Bin Khalifa University, which supported this publication. } 
	
}

\maketitle
\begin{abstract}
We consider \emph{outdoor optical access points (OAPs)}, which, enabled by recent advances in metasurface technology, have attracted growing interest. While OAPs promise high data rates and strong physical-layer security, practical deployments still expose vulnerabilities and misuse patterns that necessitate a dedicated monitoring layer—the focus of this work. 
We therefore propose a user positioning and monitoring system that infers locations from spatial intensity measurements on a photodetector (PD) array. Specifically, our hybrid approach couples an optics-informed forward model and sparse, model-based inversion with a lightweight data-driven calibration stage, yielding high accuracy at low computational cost. This design preserves the interpretability and stability of model-based reconstruction while leveraging learning to absorb residual nonidealities and device-specific distortions. Under identical hardware and training conditions (both with $5\times10^5$ samples), the hybrid method attains consistently lower mean-squared error than a generic deep-learning baseline while using substantially less training time and compute. Accuracy improves with array resolution and saturates around $60\times60$–$80\times80$, indicating a favorable accuracy–complexity trade-off for real-time deployment. The resulting position estimates can be cross-checked with real-time network logs to enable continuous monitoring, anomaly detection (e.g., potential eavesdropping), and access control in outdoor optical access networks.
\end{abstract}


\IEEEpeerreviewmaketitle


\section{Introduction}
Free-space optical (FSO) communication has emerged as a promising technology for next-generation wireless access owing to its vast unlicensed bandwidth, immunity to radio-frequency interference, and high link directionality \cite{Chow2025}. In recent years, the use of intelligent reflecting surfaces (IRSs) and modulating retroreflectors (MRRs) has facilitated the realization of lightweight and power-efficient optical access points that can extend network coverage without complex alignment requirements \cite{Chondrogiannis2025}. However, these systems remain highly sensitive to physical and environmental disturbances. The bidirectional optical path magnifies angle-of-arrival (AoA) fluctuations, atmospheric turbulence, and geometric pointing errors; even milliradian-level offsets can disrupt the link \cite{Chow2025}, \cite{Honz2025}. Consistent with this sensitivity, \cite{dabiri2022modulating} developed a comprehensive unmanned aerial vehicle (UAV)-to-ground MRR channel model incorporating tracking errors, UAV attitude fluctuations, turbulence, and range, showing that sub-milliradian pointing can interrupt the double-pass link. To improve robustness, \cite{Dabiri2023} employed MRR arrays on small UAVs and derived closed-form outage and capacity expressions under realistic turbulence and vibration, and \cite{Dabiri2025} introduced a network-coded MRR relay that demonstrated superior angular stability compared with amplify-and-forward (AF) and IRS relays.

Parallel progress in the IRS domain has targeted FSO resilience and flexibility. In \cite{Sipani2023}, an IRS-assisted FSO link under random misalignment was modeled via a unified geometric–misalignment framework capturing incidence and reflection angles; \cite{Mondal2025}  analyzed mixed FSO/active-IRS Multiple Input, Single Output (MISO)-Non-Orthogonal Multiple Access (NOMA) with imperfect channel state information (CSI), showing active IRS can mitigate turbulence; \cite{Ishida2024} optimized multilink IRS placement while accounting for turbulence and pointing errors; and \cite{Pham2024} examined optical-IRS multiuser space-division multiple access (SDMA), quantifying performance impacts of atmospheric turbulence and interference optical power.

Although FSO links are inherently secure given their narrow beams, this advantage diminishes in MRR and IRS-assisted architectures. Here, bidirectional/reflective paths widen the effective field of view (FoV), enabling in-FoV unauthorized users to access or infer channel information without disturbing the legitimate link. In \cite{dabiri2024secure}, this vulnerability was quantified for autonomous aerial vehicle (AAV)-based MRR links, deriving eavesdropping probability as a function of FoV and eavesdropper geometry, and revealing a trade-off: narrowing the FoV increases confidentiality but reduces availability under motion and turbulence.

However, none of the existing approaches provides an in-situ optical situational-awareness layer for real-time localization of unauthorized interrogators at FSO access points. We close this gap with a unified, camera-inspired monitoring framework for IRS/MRR-based outdoor access points: a wide-FoV lens focuses residual interrogator energy onto a focal-plane photodetector array to form spatial intensity maps. A geometric–radiometric forward model solved via sparse, model-based inversion, complemented by a lightweight data-driven calibration, yields interpretable, stable position estimates under turbulence and nonidealities. Cross-checking these estimates with network logs enables continuous monitoring, anomaly/eavesdropper indication, and access control—providing the missing optical situational-awareness layer for next-generation FSO access networks.

\begin{figure}
	\begin{center}
		\includegraphics[width=3.3 in]{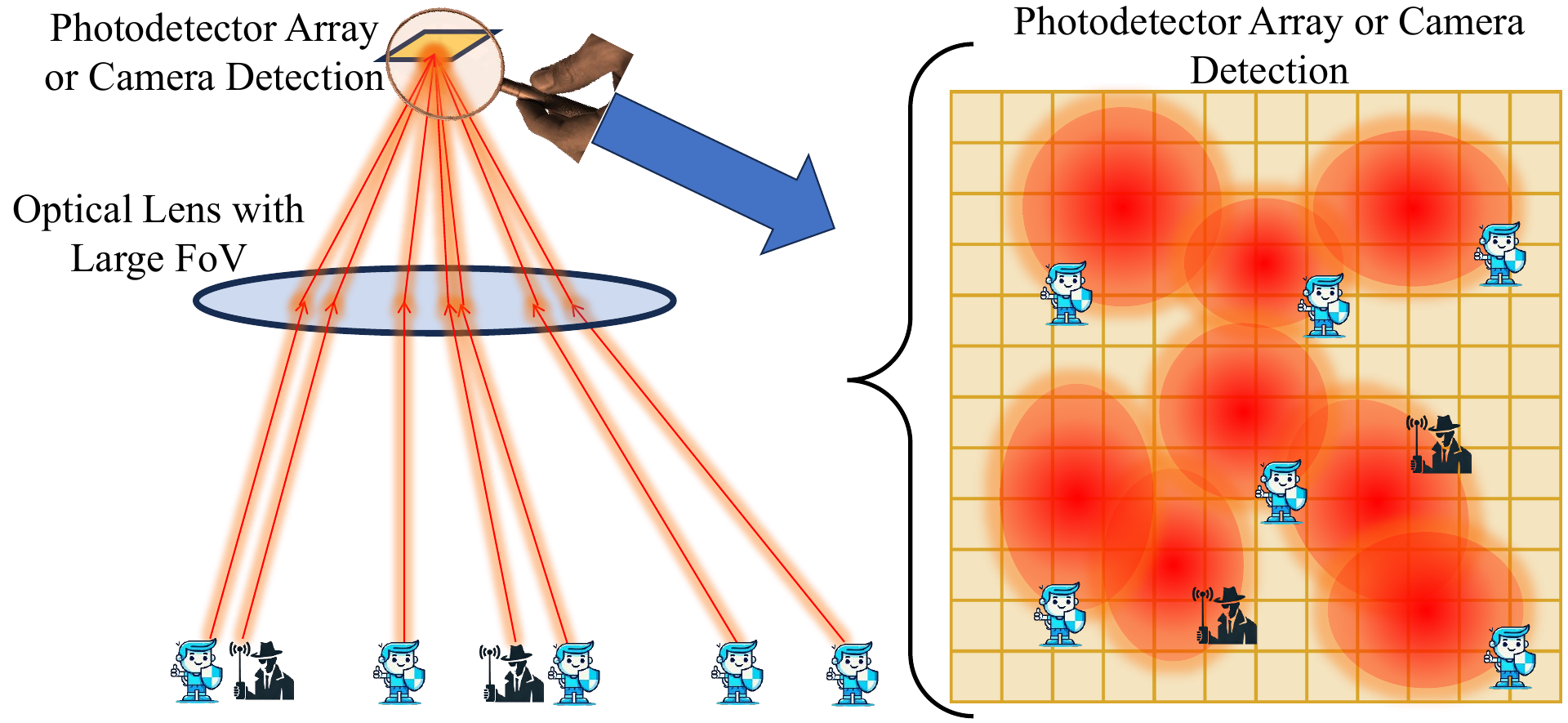}
		\caption{Illustration of the proposed monitoring system: A wide-FoV lens maps incident Gaussian beams from both legitimate users and eavesdroppers to a focal-plane photodetector array, enabling joint detection and localization.}

		\label{sm1}
	\end{center}
\end{figure}
%

\section{System Model}

\subsection{Motivation and Monitoring Framework}
In recent optical access networks based on IRSs and MRRs, accurate localization of legitimate users as well as potential eavesdroppers has become a critical requirement \cite{dabiri2025uav,dabiri2024secure}. Although such systems offer unique advantages including robustness to UAV dynamics and simplified user terminals, they are inherently exposed to a security challenge: an unauthorized interrogator located within the FoV of the IRS/MRR can illuminate the UAV and receive the modulated reflection without interfering with the legitimate link. Each ground transmitter (legitimate or eavesdropper) communicates with the UAV access point via an IRS/MRR link; however, for monitoring purposes we exploit the residual portion of their interrogator beams that naturally extends beyond the IRS/MRR aperture. At the UAV, this residual optical field impinges on the monitoring lens and can be modeled as a narrowband Gaussian beam with a planar wavefront at the lens surface due to its large spot size relative to the lens radius. 

To capture and analyze these residual beams, we propose a camera-inspired sensing module based on a focal-plane photodetector (PD) array with dimension $N_{\mathrm{ar}}=N_x\times N_y$. The array is co-located with the UAV-mounted IRS/MRR access point and placed at the focal distance of a circular lens of radius $r_a$. This monitoring plane enables two primary functions: \emph{i}) accurate estimation of legitimate users' positions through AoA and range estimation of their interrogator beams, and \emph{ii}) detection and localization of eavesdroppers attempting to exploit the link \cite{dabiri2024secure}. As depicted in Fig.~\ref{sm1}, the incident Gaussian beams are mapped to distinct focal-plane spots over the PD array, enabling joint monitoring and classification of all active transmitters within the coverage area.

\subsection{Received-Power Model for User/Eavesdropper Residual Beams}
We deploy a camera-like monitoring module (a lens co-located with the IRS/MRR access point) that focuses the portion of the optical field impinging on its lens onto a focal-plane array (Fig.~\ref{sm1}). Different user locations and ranges induce distinct AoAs and received intensities, leading to separable focal-plane patterns, albeit with a nontrivial geometric–radiometric coupling.

Let $N_u$ denote the number of legitimate users. User $i\in\{1,\ldots,N_u\}$ reports a claimed position $\mathbf p_i$ that is characterized at the UAV by its AoA
$\boldsymbol{\theta}_i=(\theta_{ix},\theta_{iy})$ and link length $L_i$; the lens plane is located at $(0,0,h_u)$, where $h_u$ is the UAV altitude. The user transmits optical power $P_i$ with divergence $\theta_{\mathrm{div},i}$ such that the Gaussian beam radius at distance $L_i$ equals $w_z$. We let the beam-axis offset relative to the lens center be $\mathbf r_{p,i}=(x_{p,i},y_{p,i})$, modeled as a random vector 
\[
x_{p,i},y_{p,i}\sim\mathcal N(\mu_p,\sigma_p^2),
\]
capturing residual boresight and platform-jitter errors \cite{dabiri2022modulating}. Under practical designs \cite{dabiri2022modulating} one typically has $w_z\gg r_a$, which justifies a plane-wave approximation across the lens aperture.

Let $\rho_i$ denote the instantaneous turbulence fading coefficient, typically modeled by a Gamma--Gamma distribution. Its probability density function (PDF) is given by \cite{Andrews2005}
\begin{align}
	f_{\rho_i}(\rho)
	= \frac{2(\alpha_i \beta_i)^{\tfrac{\alpha_i+\beta_i}{2}}}{\Gamma(\alpha_i)\,\Gamma(\beta_i)}\,
	\rho^{\tfrac{\alpha_i+\beta_i}{2}-1}\,
	K_{\alpha_i-\beta_i}\!\left(2\sqrt{\alpha_i \beta_i \rho}\right), 
	\label{eq:gamma_gamma_pdf}
\end{align}
where $\rho>0$, $\alpha_i$ and $\beta_i$ are the effective number of large- and small-scale turbulence cells, respectively, $\Gamma(\cdot)$ is the Gamma function, and $K_\nu(\cdot)$ is the modified Bessel function of the second kind.
Also, let $g_i\in(0,1]$ denote the large-scale path gain (e.g., $g_i=e^{-\kappa L_i}$ under homogeneous attenuation $\kappa$), and $\tau_\ell$ the lens transmission. The irradiance of the shifted Gaussian residual beam at the lens plane is then \cite{Dabiri2025}
\begin{align}
	\mathcal{E}_i(\mathbf r)
	&= \rho_i\,g_i\,\frac{2P_i}{\pi w_z^2}
	\exp\!\Big(-\frac{2\|\mathbf r-\mathbf r_{p,i}\|^2}{w_z^2}\Big),
	\label{eq:irradiance_field_user}
\end{align}
which becomes approximately flat over the circular lens of radius $r_a$ when $w_z\gg r_a$,
\begin{align}
	\mathcal{E}_{0,i}
	&\approx \rho_i\,g_i\,\frac{2P_i}{\pi w_z^2}
	\exp\!\Big(-\frac{2\|\mathbf r_{p,i}\|^2}{w_z^2}\Big).
	\label{eq:irradiance_flat_user}
\end{align}
Accounting for the projected lens at incidence magnitude $\theta_i=\sqrt{\theta_{ix}^2+\theta_{iy}^2}$, the total optical power collected by the lens is
\begin{align}
	P_{L,i}
	&= \tau_\ell\,\mathcal{E}_{0,i}\,(\pi r_a^2)\,\cos\theta_i \nonumber\\
	&= \tau_\ell\,\rho_i\,g_i\,
	\frac{2P_i}{\pi w_z^2}
	\exp\!\Big(-\frac{2\|\mathbf r_{p,i}\|^2}{w_z^2}\Big)\,
	(\pi r_a^2)\,\cos\theta_i.
	\label{eq:lens_power_user}
\end{align}

\section{PD Array and Lens Mapping}
We employ a focal-plane PD array of size $N_{\mathrm{ar}}=N_x\times N_y$ at the focal distance $f$ of the monitoring lens. Pixels are arranged on a Cartesian grid with pitches \(\Delta_x,\Delta_y\). Let pixel \(j\) correspond to the 2-D index \((m,n)\) and cover
\begin{align}
	x_j^{-}=x_{\min}+m\,\Delta_x,\;\;x_j^{+}=x_{\min}+(m+1)\Delta_x,\nonumber\\
	y_j^{-}=y_{\min}+n\,\Delta_y,\;\;y_j^{+}=y_{\min}+(n+1)\Delta_y.
	\label{eq:pixel_bounds_camera}
\end{align}

\subsubsection{Angle-to-focal mapping}
A plane (or locally planar) wave impinging on the lens with incident angles \(\boldsymbol{\theta}_i=(\theta_{ix},\theta_{iy})\) is focused to a point on the focal plane located at \cite{Goodman2005}
\begin{align}
	x_i=f\,\tan\theta_{ix},\qquad y_i=f\,\tan\theta_{iy}.
	\label{eq:f_tan}
\end{align}

\subsubsection{Focal-plane irradiance of one transmitter}
Let \(P_{L,i}\) be the optical power collected by the lens from user \(i\) in \eqref{eq:lens_power_user}. We model the focal-plane spot by a circular Gaussian with variance \(\sigma_{\mathrm{core}}^2\) at \(z=f\), where \(\sigma_{\mathrm{core}}^2=\sigma_{\mathrm{diff},0}^2+\sigma_{\mathrm{sens}}^2\) aggregates the diffraction-limited core at focus and sensor blur \cite{safi2025cubesat}:
\begin{align}
	I_i(x,y)=\frac{\eta_s P_{L,i}}{2\pi\sigma_{\mathrm{core}}^2}\,
	\exp\!\left(-\frac{(x-x_i)^2+(y-y_i)^2}{2\sigma_{\mathrm{core}}^2}\right),
	\label{eq:spot_intensity}
\end{align}
where $D=2r_a$ is the lens aperture diameter, and \(\eta_s\) is the optical-to-sensor efficiency.

\subsection{Geometry and Angular Coverage}
The sensor plane is shifted from the focal distance to a general axial position $z\in(0,f]$ measured from the lens. The mapping between the incident angles $\boldsymbol{\theta}=(\theta_x,\theta_y)$ and the sensor-plane coordinates is
\begin{align}
	x(z)=z\tan\theta_x,\qquad y(z)=z\tan\theta_y.
	\label{eq:map_general}
\end{align}
Let $X_{\max},Y_{\max}$ denote the sensor half-widths. The corresponding angular limits are
\begin{align}
	|\theta_x|\le \arctan\!\Big(\frac{X_{\max}}{z}\Big),\qquad
	|\theta_y|\le \arctan\!\Big(\frac{Y_{\max}}{z}\Big),
	\label{eq:fov_limits}
\end{align}
implying that smaller $z$ increases the admissible field angles and reduces off-sensor beam loss.

\subsection{Defocus and Diffraction Modeling}
When the sensor plane is displaced from the focal distance, the beam cross-section on the sensor becomes a scaled projection of the pupil. The geometric disk diameter reduces approximately linearly from the lens aperture to the focal point as \cite{Goodman2005}
\begin{align}
	d_{\mathrm{geo}}(z)=2r_a\Big|1-\frac{z}{f}\Big|,
	\label{eq:dgeo}
\end{align}
where $z$ is the sensor distance from the lens and $r_a$ is the lens radius. 
The Gaussian-equivalent variance of this geometric defocus pattern is \cite{Goodman2005}
\begin{align} 
	\sigma_{\mathrm{def}}^2(z)=\frac{d_{\mathrm{geo}}^2(z)}{16}=\frac{D^2}{16f^2}\Delta z^2,
	\label{eq:sigma_def}
\end{align}
where $\Delta z=f-z$ and $D=2r_a$.
In addition to geometric blur, diffraction imposes a lower bound on the achievable spot size. The diffraction-limited core formed by the lens aperture at focus can be represented by a Gaussian spot with variance \(\sigma_{\mathrm{diff},0}^2\!\propto\!(\lambda f/D)^2\) at $z=f$. 
When the sensor plane is moved to a general position $z$, the diffraction component scales with the magnification factor $(z/f)$, giving \cite{Goodman2005}
\begin{align}
	\sigma_{\mathrm{diff}}^2(z)=\Big(\frac{z}{f}\Big)^2\sigma_{\mathrm{diff},0}^2.
	\label{eq:sigma_diff}
\end{align}
The effective on-sensor variance that jointly accounts for diffraction, defocus, and sensor blur is therefore
\begin{align}
	\sigma_{\mathrm{eff}}^2(z)=\sigma_{\mathrm{def}}^2(z)+\sigma_{\mathrm{diff}}^2(z)+\sigma_{\mathrm{sens}}^2.
	\label{eq:sigma_eff}
\end{align}

\subsection{Pixel Signal and Crosstalk}
Let $y_{j,e}$ denote the (noiseless) contribution of an eavesdropper $e$ to pixel $j$. 
Using the collected power $P_{L,i}$ in \eqref{eq:lens_power_user} and the mapping \eqref{eq:map_general}, the focal-plane irradiance of user $i$ is \cite{Dabiri2025}
\begin{align}
	&I_i(x,y;z)= \nonumber \\
	&\frac{\eta_s P_{L,i}}{2\pi\sigma_{\mathrm{eff}}^2(z)}
	\exp\!\Bigg(-\frac{(x-x_i(z))^2+(y-y_i(z))^2}{2\sigma_{\mathrm{eff}}^2(z)}\Bigg),
	\label{eq:spot_intensity_z}
\end{align}
where $x_i(z)=z\tan\theta_{ix}$ and $y_i(z)=z\tan\theta_{iy}$.
For pixel $j$ defined in \eqref{eq:pixel_bounds_camera}, the contribution of user $i$ is
\begin{align}
	y_{j,i}(z)
	&=T_{\mathrm{int}}\frac{\eta_s P_{L,i}}{2\pi\sigma_{\mathrm{eff}}^2(z)}
	\Big[\Phi\!\Big(\tfrac{x_j^+-x_i(z)}{\sigma_{\mathrm{eff}}(z)}\Big)-\Phi\!\Big(\tfrac{x_j^--x_i(z)}{\sigma_{\mathrm{eff}}(z)}\Big)\Big]\nonumber\\
	&\quad\times
	\Big[\Phi\!\Big(\tfrac{y_j^+-y_i(z)}{\sigma_{\mathrm{eff}}(z)}\Big)-\Phi\!\Big(\tfrac{y_j^--y_i(z)}{\sigma_{\mathrm{eff}}(z)}\Big)\Big]
	\label{eq:yji_z}
\end{align}
where \(\Phi(\cdot)\) is the standard normal cumulative distribution function (CDF),
$D=2r_a$ is the lens aperture diameter, and \(\eta_s\) is the optical-to-sensor efficiency

\subsubsection{Output of pixel $j$}
The measured output of pixel $j$ over an integration time $T_{\mathrm{int}}$ is then
\begin{align}
	y_j 
	&= \sum_{i=1}^{N_u} y_{j,i} \;+\; \sum_{e=1}^{N_e} y_{j,e} + n_j,
	\qquad j\in\{1,\ldots,N_{\mathrm{ar}}\},
	\label{eq:meas_total}
\end{align}
where $n_j\sim\mathcal{N}(0,N_0)$ is thermal noise.

\begin{figure}
	\centering
	\subfloat[] {\includegraphics[width=1.65 in]{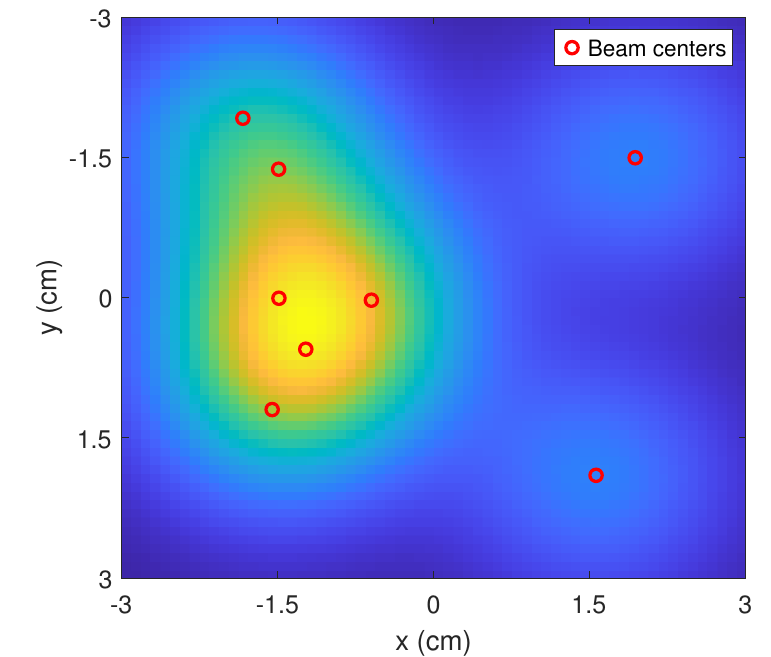}
		\label{cb1}
	}
	\hfill
	\subfloat[] {\includegraphics[width=1.65 in]{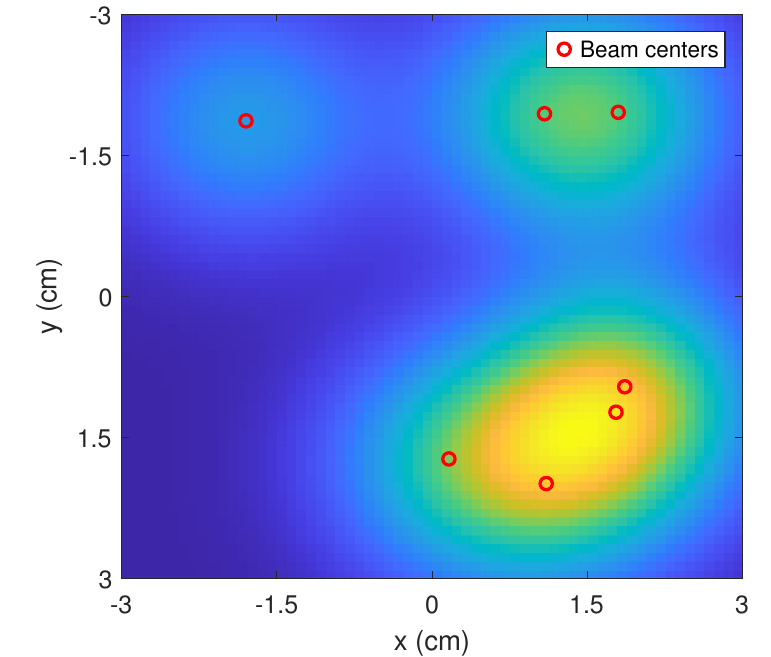}
		\label{cb2}
	}	
	\caption{Two random examples of focal-plane irradiance patterns produced by simultaneous user transmissions under random incidence angles, atmospheric turbulence, and pointing jitter (a) for $N_u=8$ users; and (b) for $N_u=7$ users.}
	\label{cb}
\end{figure}

\section{User Localization via Hybrid Deep Learning and Optimal Beam Matching}
\label{sec:localization}

\subsection{Objective}
The focal-plane detector yields an unordered set of spot centroids produced by multiple concurrent transmitters whose irradiance patterns may overlap. Because centroids carry no identity, localization requires a permutation-invariant correspondence between the detected peaks and the physical sources before any geometric inversion is applied. For instance, in two random realizations shown in Fig.~\ref{cb}, the spatial information becomes highly entangled due to random incidence angles, turbulence-induced beam wander, and pointing jitter, which cause several focal-plane spots to merge or shift, making direct user-to-beam association ambiguous.
Therefore our target is to seek a permutation-invariant procedure that (i) extracts reliable centroids from the irradiance and (ii) recovers a one-to-one correspondence between detected peaks and physical sources, after which geometric inversion yields user (or eavesdropper) locations.

\subsection{Observation Model}
Let $\mathcal{R}=\{\boldsymbol{r}_i\}_{i=1}^{N}$ denote the unknown spot centroids on the sensor plane, with $\boldsymbol{r}_i=[x_i(z),y_i(z)]^{\!\top}$ given by the established mapping \eqref{eq:map_general}. The focal-plane irradiance is a superposition of Gaussian-like spots whose effective spread obeys \eqref{eq:sigma_eff}. The detector (learned or analytic) provides an \emph{unordered} set of candidate centroids $\widehat{\mathcal{R}}=\{\widehat{\boldsymbol{r}}_j\}_{j=1}^{N}$. When spots interpenetrate (cf. Fig.~\ref{cb}), local maxima can shift or merge, so identities are not observable.

The mapping \eqref{eq:map_general} converts AoA into focal-plane coordinates, while \eqref{eq:sigma_eff} ties optical blur (diffraction, defocus, sensor) to centroid uncertainty. Hence any estimator must be (i) \emph{permutation-invariant} (identities are not encoded in the optics) and (ii) \emph{aware of blur statistics} (heterogeneous $\sigma_{\mathrm{eff}}$ yields heteroscedastic centroid noise). These two requirements drive the proposed hybrid design.

\subsection{Heatmap Inference (Deep Learning)}
A convolutional network maps the irradiance to a single-channel heatmap $\widehat{H}(x,y)\in[0,1]$ whose peaks indicate centroids. The target is a permutation-invariant superposition of isotropic Gaussians centered at $\mathcal{R}$, clipped to $[0,1]$, with standard deviation chosen to match optics, e.g., $\sigma_{\text{tgt}}\!\approx\!\sigma_{\mathrm{eff}}$.

A shallow encoder–decoder with $3\times3$ convolutions, batch normalization (BN), and rectified linear unit (ReLU) provides translation-equivariant feature extraction aligned with the near-stationary point spread function (PSF) structure; a $1\times1$ head with sigmoid produces $\widehat{H}$. Training minimizes $\|\widehat{H}-H_{\text{tgt}}\|_2^2$, which encourages energy concentration at true centroids without enforcing identities. This design is physically consistent: the convolutional neural network (CNN) learns to disentangle overlapping irradiance patterns while the target blur mirrors \eqref{eq:sigma_eff}, ensuring matched bias–variance \cite{Goodfellow2016}.

\subsection{Peak Extraction}
Local maxima are extracted from $\widehat{H}$ using non-maximum suppression (NMS) with radius $r_{\text{NMS}}\!\propto\!\sigma_{\mathrm{eff}}$. This suppresses adjacent spurious peaks yet preserves dominant modes when spots partially merge. The result is the unordered set $\widehat{\mathcal{R}}$ of $N$ centroids.

\subsection{Optimal Correspondence (Assignment)}
Define the cost matrix \cite{Goodfellow2016}:
\begin{align}
	D_{ij}\triangleq \|\boldsymbol{r}_i-\widehat{\boldsymbol{r}}_j\|_2,\qquad i,j=1,\ldots,N.
	\label{eq:dist_euclid_local}
\end{align}
Recover the identity-free correspondence via the linear assignment problem (LAP):
\begin{align}
	\min_{\boldsymbol{\Pi}\in\{0,1\}^{N\times N}}
	\sum_{i=1}^{N}\sum_{j=1}^{N}\Pi_{ij}\,D_{ij}
	\;\;\text{s.t.}\;\;
	\boldsymbol{\Pi}\mathbf{1}=\mathbf{1},\;
	\boldsymbol{\Pi}^{\!\top}\mathbf{1}=\mathbf{1},
	\label{eq:LAP_local}
\end{align}
whose optimizer $\boldsymbol{\Pi}^\star$ is a permutation matrix giving a one-to-one mapping between detected and physical centroids. Polynomial-time solvers (e.g., Kuhn–Munkres) deliver the global optimum and avoid greedy failures under interpenetrating spots.

\begin{figure*}
	\centering
	\subfloat[] {\includegraphics[width=2.30 in]{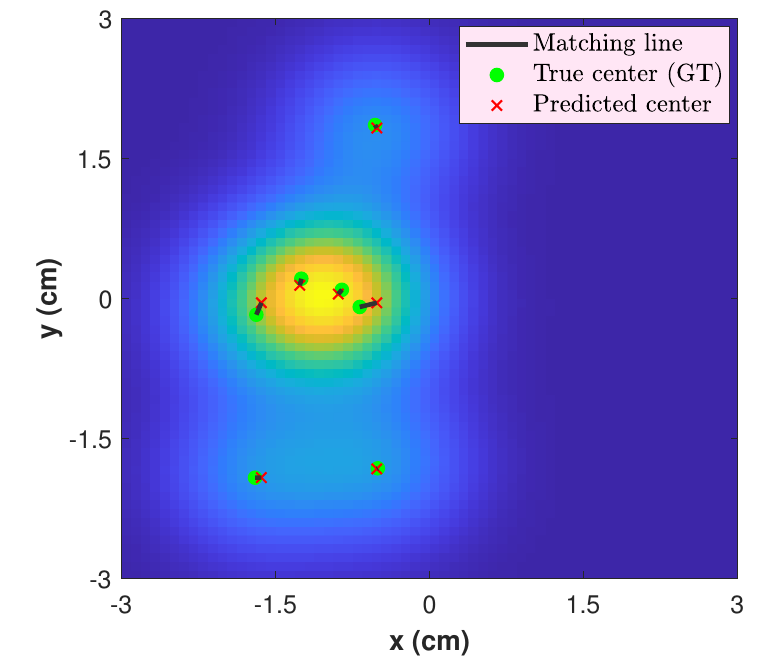}
		\label{cf1}
	}
	\hfill
	\subfloat[] {\includegraphics[width=2.30 in]{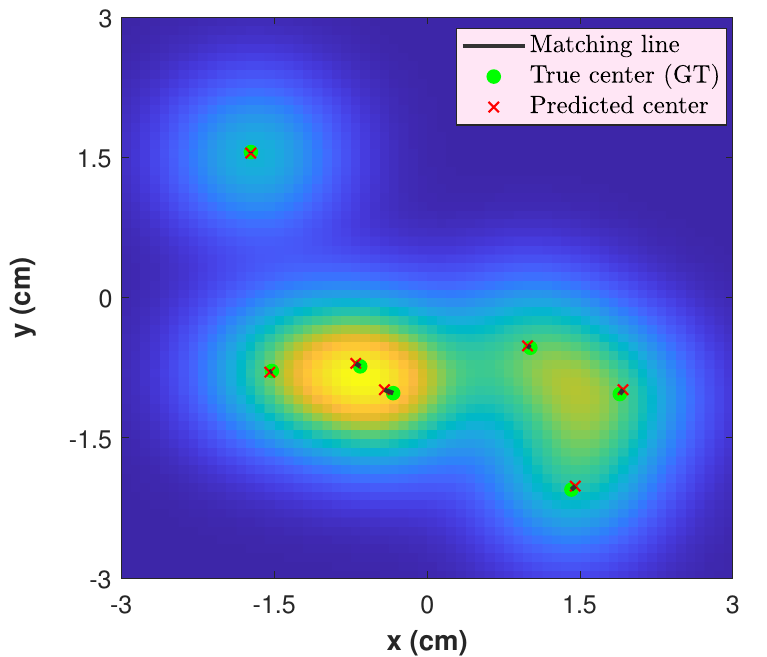}
		\label{cf2}
	}	
	\hfill
	\subfloat[] {\includegraphics[width=2.30 in]{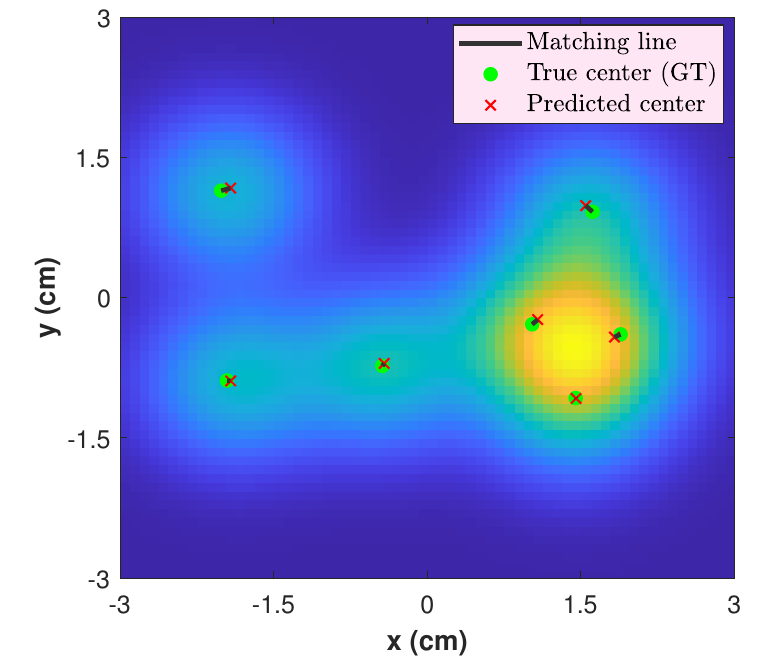}
		\label{cf3}
	}
	\caption{Three random test cases of user localization with a 60$\times$60 photodetector array. Heatmaps show received optical intensity on the array surface; green dots denote ground-truth user positions on the array plane, and red crosses are the estimated positions.}
	\label{cf}
\end{figure*}

\subsubsection{Statistical Optimality and Physical Weighting}
If centroid errors are Gaussian with per-source covariance $\boldsymbol{\Sigma}_i$ (induced by \eqref{eq:sigma_eff}), the negative log-likelihood of an assignment is proportional to \cite{Goodfellow2016}:
\begin{align}
	\sum_{i,j}\Pi_{ij}\,
	\big\|\boldsymbol{r}_i-\widehat{\boldsymbol{r}}_j\big\|_{\boldsymbol{\Sigma}_i^{-1}}^{2}.
	\label{eq:mahal_cost}
\end{align}
Replacing \eqref{eq:dist_euclid_local} by the Mahalanobis cost
$D_{ij}^{(\mathrm{w})}=\|\boldsymbol{r}_i-\widehat{\boldsymbol{r}}_j\|_{\boldsymbol{\Sigma}_i^{-1}}$
yields a maximum likelihood (ML) (and maximum a posteriori (MAP) under uniform priors) correspondence that is \emph{physics-aware}: beams with larger $\sigma_{\mathrm{eff}}$ naturally receive lower confidence. In the isotropic homogeneous case, Euclidean costs remain ML-optimal.

\subsubsection{Outliers and Cardinality Mismatch}
With spurious or missed peaks, augment the smaller side with $M$ dummy nodes and penalty $c_{\max}$ to form a rectangular LAP; matches to dummies declare unassigned items and cap false associations while preserving global optimality.

\subsection{Inverse Geometric Reconstruction}
Ordering detections by $\boldsymbol{\Pi}^\star$ as $\{\widehat{\boldsymbol{r}}_i\}_{i=1}^{N}$, the incident angles and lateral coordinates follow from the inverse of \eqref{eq:map_general}:
\begin{align}
	\widehat{\theta}_{ix}=\arctan\!\big(\widehat{x}_i/z\big),
	\widehat{\theta}_{iy}=\arctan\!\big(\widehat{y}_i/z\big),
	\widehat{\boldsymbol{p}}_i
	= h_u\!\begin{bmatrix}\tan\widehat{\theta}_{ix}\\ \tan\widehat{\theta}_{iy}\end{bmatrix}.
	\label{eq:inverse_geo_final}
\end{align}
These relations yield user (or eavesdropper) locations once correspondence is resolved.

\subsection{Why This Hybrid is Physically Justified}
The optics impose two constraints: (i) identities are not encoded in the irradiance (only geometry is), and (ii) centroid uncertainty is governed by \eqref{eq:sigma_eff}. The deep learning (DL) heatmap converts a complex superposition into a peak-enhanced, permutation-agnostic representation, while the LAP enforces the one-to-one geometric pairing that the physics demands. Weighting via \eqref{eq:mahal_cost} injects $\sigma_{\mathrm{eff}}$ into the estimator, making it robust under heterogeneous blur consistent with Fig.~\ref{cb}.

\subsection{Complexity and Guarantees}
CNN inference scales with image size and depth; NMS is near-linear in pixels; the LAP scales as $\mathcal{O}(N^{3})$ and is small for moderate numbers of active sources. The correspondence is globally optimal for the chosen metric, and the overall pipeline is permutation-invariant by construction and consistent with the underlying optical model.
\section{Simulation Setup and Benchmarking Framework}
\label{sec:simulation}

\subsection{Simulation Parameters and Environment}
The performance of the proposed hybrid monitoring framework was evaluated through a comprehensive Monte Carlo simulation. The simulation emulates the UAV-mounted optical sensing system described in Sections~II–III, generating synthetic focal-plane irradiance data under realistic turbulence, noise, and pointing-jitter conditions. 
In each Monte Carlo realization, the total number of transmitters is \(N = N_u + N_e\), where \(N \in \{5,6,7,8,9\}\), and the monitoring system jointly estimates \(N\), their spatial coordinates, and potential eavesdropper locations.

Simulation parameters are selected as typical values reported in the literature:  
\(h_u = 300~\mathrm{m},\; L_{\mathrm{area}} = 250{\times}250~\mathrm{m}^2,\; P_t = 1~\mathrm{W},\;
w_z = 0.6~\mathrm{m},\; \sigma_p = 10~\mathrm{cm},\;
r_a = 0.05~\mathrm{m},\; \tau_\ell = 0.9,\;
\eta_s = 0.9,\; T_{\mathrm{int}} = 1~\mathrm{s},\;
N_0 = 10^{-12}~\mathrm{W},\; (\alpha,\beta) = (4,2),\; \sigma_{\mathrm{eff}} =8 ~\mathrm{mm},\;
\kappa = 0.001,\; N \in \{5,6,7,8,9\}.\)
The channel coherence time is \(T_{\mathrm{coh}} = 10~\mathrm{ms}\), 
yielding approximately \(T_{\mathrm{int}} / T_{\mathrm{coh}} = 100\) statistically independent fading realizations within each observation period.
By default, \(N_x = N_y = 64\), varied between \(10\times10\) and \(100\times100\).  
The sensor area is \(3\times3~\mathrm{cm}^2\), giving pixel pitch \(\Delta_x = \Delta_y = 3~\mathrm{cm}/N_x\).  
The beam divergence is \( \sigma_{\mathrm{eff}} =8 ~\mathrm{mm}\) by default, varied within \(1\!-\!10~\mathrm{mm}\) for performance analysis.

\subsection{Proposed Network Configuration}
The proposed network follows a single-heatmap structure with mean-squared-error (MSE) training and NMS for inference.  
The input is a normalized focal-plane irradiance map of size \(N_y\times N_x\times1\), and the target is a clipped Gaussian heatmap with peaks equal to one.  
The model consists of four convolutional layers with batch normalization and ReLU activation, followed by a sigmoid output layer producing \(\widehat{H}(x,y)\in[0,1]\).  
During inference, the top-\(N\) peaks are selected using NMS, and their positions are matched to the ground truth via optimal assignment as in~\eqref{eq:LAP_local}.  
This configuration enables permutation-invariant detection and localization of multiple transmitters under realistic turbulence and noise conditions.

\subsection{Benchmark Deep-Learning Model}
For comparison, a fully data-driven end-to-end network without optical priors is implemented.  
It adopts a U-Net–like encoder–decoder with four downsampling and four upsampling stages using $3{\times}3$ convolutions, batch normalization, and LeakyReLU.  
Two output branches generate a sigmoid heatmap and coordinate maps, trained with the composite loss
\begin{align}
	\mathcal{L} = 
	\lambda_1 \| \widehat{\mathbf{C}} - \mathbf{C}_{\mathrm{true}} \|_2^2 
	+ \lambda_2\, \mathcal{L}_{\mathrm{BCE}}(\widehat{H}, H_{\mathrm{true}}),
\end{align}
where $(\lambda_1,\lambda_2)=(1,0.5)$.  
This model has about $3.5$~M parameters and serves as a black-box baseline for evaluating the proposed hybrid method under varying and unknown numbers of transmitters~$N$.

\begin{table}[t]
	\centering
	\caption{Comparison of Reconstruction Accuracy for Different Photodetector Array Sizes}
	\begin{tabular}{ccc}
		\hline
		Photodetector Array Size & MSE (Proposed) & MSE (Deep Learning) \\
		\hline
		10$\times$10  & 15.3 m$^2$  & 19.7 m$^2$ \\
		20$\times$20  & 8.2 m$^2$   & 12.5 m$^2$  \\
		40$\times$40  & 2.6 m$^2$   & 5.4 m$^2$  \\
		60$\times$60  & 0.9 m$^2$   & 2.3 m$^2$  \\
		80$\times$80  & 0.84 m$^2$  & 1.7 m$^2$ \\
		100$\times$100 & 0.81 m$^2$ & 1.3 m$^2$ \\
		\hline
	\end{tabular}
	\vspace{2mm}
	\begin{flushleft}
		\footnotesize{Both methods were trained using $5\times10^5$ samples. The training time of the deep learning network was approximately twice that of the proposed system under the same hardware conditions.}
	\end{flushleft}
	\label{tab:mse_comparison}
\end{table}

\subsection{Performance Analysis and Discussion}
Table~\ref{tab:mse_comparison} presents the positioning accuracy obtained for different photodetector array sizes in the user-location monitoring setup. As the number of photodetector elements increases, the system receives a denser spatial representation of the captured optical field, resulting in lower MSE values and improved accuracy in user position reconstruction.
From a computational perspective, increasing the number of photodetector pixels is equivalent to increasing the dimensionality of the input feature space. Each additional sensing element introduces a new input variable, which linearly increases the size of the data vector and consequently the number of operations required per iteration in both the proposed and deep learning frameworks. While this extension improves accuracy, it also enlarges the model’s memory footprint and training time.

In the proposed method, the computational complexity scales approximately linearly with the number of pixels, since the model utilizes an optimized mapping and sparse transformation structure. In contrast, the deep learning-based method exhibits higher-order complexity due to its dense neural network architecture, large parameter space, and iterative backpropagation steps. As a result, the deep learning model requires roughly twice the training time under identical computational settings, even though both systems were trained on $5\times10^5$ samples.  Overall, the proposed approach not only achieves lower MSE values across all configurations but also requires substantially less computational time and resources compared to the deep learning-based method. This demonstrates its superior balance between positioning accuracy and computational efficiency, making it more suitable for real-time user monitoring systems.

Another important observation is that beyond array configurations of 60$\times$60 or 80$\times$80, the reduction in MSE becomes negligible, indicating a saturation in accuracy improvement. This suggests that further increasing the photodetector resolution primarily increases computational load without providing a proportional benefit in performance. Therefore, an optimal design should consider both reconstruction accuracy and computational efficiency, especially for real-time user monitoring applications where low latency and resource constraints are critical.

Figure~\ref{cf} illustrates three random examples of user positioning results obtained using a 60$\times$60 photodetector array. In each subfigure, the heatmap represents the spatial intensity distribution of the received optical signal across the photodetector surface, where brighter regions indicate stronger received power. The green dots denote the true user positions, while the red crosses represent the estimated user positions obtained by the proposed positioning model.
The examples demonstrate that the proposed method can accurately predict user locations in most scenarios, with a small average deviation between the predicted and true centers. However, as shown in Fig.~\ref{cf1}, when multiple users are positioned very close to each other on the photodetector plane, the strong overlap between their received optical signatures can lead to higher localization errors due to mutual interference. This effect is particularly noticeable when the users’ optical responses partially overlap, reducing the separability of their spatial features.
These observations emphasize the robustness of the proposed system in typical multi-user monitoring scenarios while highlighting its sensitivity to dense spatial configurations. Incorporating advanced separation techniques or adaptive filtering can further mitigate such interference effects in future implementations.

Furthermore, the estimated user-positioning information can be directly compared with real-time feedback or ground-truth data collected from optical access networks. Such correlation enables continuous user monitoring, anomaly detection, and the identification of suspicious behaviors such as potential eavesdropping or irregular access patterns. This capability also provides a practical foundation for intelligent user management and access control in next-generation optical wireless networks, where spatial awareness and security are critical for system reliability and performance.



\bibliographystyle{IEEEtran}
\bibliography{IEEEabrv,myref}

\begin{thebibliography}{10}
\providecommand{\url}[1]{#1}
\csname url@samestyle\endcsname
\providecommand{\newblock}{\relax}
\providecommand{\bibinfo}[2]{#2}
\providecommand{\BIBentrySTDinterwordspacing}{\spaceskip=0pt\relax}
\providecommand{\BIBentryALTinterwordstretchfactor}{4}
\providecommand{\BIBentryALTinterwordspacing}{\spaceskip=\fontdimen2\font plus
\BIBentryALTinterwordstretchfactor\fontdimen3\font minus
  \fontdimen4\font\relax}
\providecommand{\BIBforeignlanguage}[2]{{%
\expandafter\ifx\csname l@#1\endcsname\relax
\typeout{** WARNING: IEEEtran.bst: No hyphenation pattern has been}%
\typeout{** loaded for the language `#1'. Using the pattern for}%
\typeout{** the default language instead.}%
\else
\language=\csname l@#1\endcsname
\fi
#2}}
\providecommand{\BIBdecl}{\relax}
\BIBdecl

\bibitem{Chow2025}
C.-W. Chow, ``Optical wireless communication – recent progresses and future
  perspectives,'' \emph{Journal of Lightwave Technology}, 2025.

\bibitem{Chondrogiannis2025}
G.~D. Chondrogiannis, A.~P. Chrysologou, A.-A.~A. Boulogeorgos, N.~D.
  Chatzidiamantis, and H.~Haas, ``Optical ris-enabled multiple access
  communications,'' \emph{IEEE Transactions on Green Communications and
  Networking}, 2025.

\bibitem{Honz2025}
F.~Honz and B.~Schrenk, ``Turbulence-resilient reflective fi-wi-fi bridge for
  terrestrial free-space optical data links,'' \emph{Journal of Lightwave
  Technology}, 2025.

\bibitem{dabiri2022modulating}
M.~T. Dabiri, M.~Rezaee, L.~Mohammadi, F.~Javaherian, V.~Yazdanian, M.~O.
  Hasna, and M.~Uysal, ``{Modulating retroreflector based free space optical
  link for UAV-to-ground communications},'' \emph{IEEE Transactions on Wireless
  Communications}, vol.~21, no.~10, pp. 8631--8645, 2022.

\bibitem{Dabiri2023}
M.~T. Dabiri and M.~Hasna, ``Performance analysis of modulating retroreflector
  array for uav-based fso links,'' \emph{IEEE Communications Letters}, vol.~27,
  no.~12, pp. 3280--3284, Dec. 2023.

\bibitem{Dabiri2025}
------, ``A novel mrr-uav-based relay with optical network coding: A
  comparative study with optical irs and conventional uav relaying,''
  \emph{IEEE Journal on Selected Areas in Communications}, vol.~43, no.~5, pp.
  1607--1620, May 2025.

\bibitem{Sipani2023}
J.~Sipani, P.~Sharda, and M.~R. Bhatnagar, ``Modeling and design of
  irs-assisted fso system under random misalignment,'' \emph{IEEE Photonics
  Journal}, vol.~15, no.~4, pp. 1--13, Aug. 2023, art. no. 7303113.

\bibitem{Mondal2025}
S.~Mondal, K.~Singh, C.-P. Li, and Z.~Ding, ``Mixed fso/active irs-aided miso
  noma communication with imperfect csi and sic,'' \emph{IEEE Internet of
  Things Journal}, vol.~12, no.~1, pp. 623--636, Jan. 2025.

\bibitem{Ishida2024}
T.~Ishida, C.~B. Naila, H.~Okada, and M.~Katayama, ``Performance analysis of
  irs-assisted multi-link fso system under pointing errors,'' \emph{IEEE
  Photonics Journal}, vol.~16, no.~4, pp. 1--10, Aug. 2024, art. no. 7302510.

\bibitem{Pham2024}
P.~D. Pham, C.~K.~P. Nguyen, H.~D. Le, H.~T.~T. Pham, T.~V. Nguyen, and N.~T.
  Dang, ``Optical intelligent reflecting surface-assisted multiple users over
  turbulence channels,'' in \emph{Proceedings of the 2024 RIVF International
  Conference on Computing and Communication Technologies (RIVF)}, Danang,
  Vietnam, 2024, pp. 45--50.

\bibitem{dabiri2024secure}
M.~T. Dabiri, M.~Hasna, S.~Althunibat, and K.~Qaraqe, ``{How Secure are
  AAV-Based FSO Links With Modulating Retroreflectors?}'' \emph{IEEE Wireless
  Communications Letters}, vol.~14, no.~3, pp. 606--610, 2025.

\bibitem{dabiri2025uav}
------, ``{UAV-BASED Dynamic FSO Access Networks: Technological Comparison,
  Design Considerations, and Future Directions},'' \emph{IEEE Wireless
  Communications}, vol.~32, no.~2, pp. 247--253, 2025.

\bibitem{Andrews2005}
L.~C. Andrews and R.~L. Phillips, \emph{Laser Beam Propagation through Random
  Media}, 2nd~ed.\hskip 1em plus 0.5em minus 0.4em\relax Bellingham, WA, USA:
  SPIE Press, 2005, print ISBN: 9781510643703.

\bibitem{Goodman2005}
J.~W. Goodman, \emph{Introduction to Fourier Optics}, 3rd~ed.\hskip 1em plus
  0.5em minus 0.4em\relax Greenwood Village, CO, USA: Roberts and Company
  Publishers, 2005.

\bibitem{safi2025cubesat}
H.~Safi, M.~T. Dabiri, J.~Cheng, I.~Tavakkolnia, and H.~Haas,
  ``{CubeSat-Enabled Free-Space Optics: Joint Data Communication and Fine Beam
  Tracking},'' \emph{IEEE Transactions on Vehicular Technology}, 2025.

\bibitem{Goodfellow2016}
I.~Goodfellow, Y.~Bengio, and A.~Courville, \emph{Deep Learning}.\hskip 1em
  plus 0.5em minus 0.4em\relax Cambridge, MA, USA: MIT Press, 2016.

\end{thebibliography}

\end{document}